\documentclass[%
 aip,
 amsmath,amssymb,
 reprint,%
]{revtex4-1}
\usepackage{xcolor}
\usepackage{multirow}
\usepackage{graphicx}
\usepackage{dcolumn}
\usepackage{bm}

\usepackage[utf8]{inputenc}
\usepackage[T1]{fontenc}
\usepackage{mathptmx}
\usepackage{etoolbox}
\usepackage{xspace} 
\usepackage{booktabs}
\usepackage{tikz}
\definecolor{shsPurple}{RGB}{191,0,255}

\DeclareRobustCommand{\blackline}{\tikz{\draw[black, thick](0,0)--(0.6,0);}}
\DeclareRobustCommand{\bluedash}{\tikz{\draw[blue, dashed, thick](0,0)--(0.6,0);}}
\DeclareRobustCommand{\greendot}{\tikz{\draw[green!60!black, dotted, thick](0,0)--(0.6,0);}}
\DeclareRobustCommand{\purpledashdot}{\tikz{\draw[shsPurple, dash dot, thick](0,0)--(0.6,0);}}

\makeatletter
\def\@email#1#2{%
 \endgroup
 \patchcmd{\titleblock@produce}
  {\frontmatter@RRAPformat}
  {\frontmatter@RRAPformat{\produce@RRAP{*#1\href{mailto:#2}{#2}}}\frontmatter@RRAPformat}
  {}{}
}%
\makeatother
\begin{document}

\preprint{AIP/123-QED}

\title{Uncertainty Quantification of Drag Reduction over Superhydrophobic Surfaces by Unified Parameterizing Structure Spacing}

\author{Byeong-Cheon Kim}
\affiliation{School of Mechanical Engineering, University of Ulsan, Ulsan}

\author{Kyoungsik Chang}%
\altaffiliation{School of Mechanical Engineering, University of Ulsan, Ulsan}
\email{kschang76@ulsan.ac.kr}
 
\author{Sang-Wook Lee}
\affiliation{School of Mechanical Engineering, University of Ulsan, Ulsan}%
\author{Hoai Thanh Nguyen}
\affiliation{School of Mechanical Engineering, University of Ulsan, Ulsan}%
\author{Eun Seok Oh}
\affiliation{School of Mechanical Engineering, University of Ulsan, Ulsan}%

\author{Jaiyoung Ryu}
\affiliation{ 
School of Mechanical Engineering, Korea University, Seoul
}%
\author{Minjae Kim}
\affiliation{ 
5th R$\&$D Institute, Agency for Defense Development, Changwon
}%
\author{Jaemoon Yoon}
\affiliation{ 
5th R$\&$D Institute, Agency for Defense Development, Changwon
}%

\date{\today}

\begin{abstract}
Superhydrophobic surfaces (SHS) have demonstrated significant potential in reducing turbulent drag by introducing slip conditions through micro-structured geometries. While previous studies have examined individual SHS configurations such as post-type, ridge-type, and transverse ridge-type surfaces, a unified analysis that connects these patterns through geometric parameterization remains limited. In this study, we propose a systematic framework to explore the drag reduction characteristics by varying the streamwise and spanwise spacing ($d_1, d_2$) of post-type patterns, effectively encompassing a range of SHS geometries. High-fidelity direct numerical simulations (DNS) were performed using NekRS, a GPU-accelerated spectral element solver, to resolve incompressible turbulent channel flows over these SHSs. To account for variability in the geometric parameters and quantify their influence, we construct a surrogate model based on polynomial chaos expansion (PCE) using Latin hypercube sampling (LHS) method. The resulting model enables efficient uncertainty quantification (UQ) and sensitivity analysis, revealing the relative importance of $d_1$ and $d_2$ in drag reduction performance. This unified UQ framework provides both predictive capability and design guidance for optimizing SHS configurations under uncertain geometric conditions.

\end{abstract}

\maketitle

\section{\label{sec:level1}Introduction}

Superhydrophobic surface (SHS) is characterized by micro- or nano-scale surface textures that induce extreme water repellency. Inspired by natural surfaces such as lotus leaves, SHS can effectively modify the fluid–solid boundary condition by introducing slip, thereby reducing drag in both laminar and turbulent flows\cite{rothstein2010slip,lee2016superhydrophobic,park2021superhydrophobic}. As a result, they have attracted significant attention in recent years due to their remarkable ability to reduce frictional drag in turbulent flows, a property beneficial for various engineering applications, including microfluidics and maritime vessels\cite{luo2024recent,chen2023open,wang2023review}. This drag reduction is primarily attributed to the formation of stable air pockets trapped within the surface textures, which create an air-water interface, minimizing direct fluid-solid interactions. The effectiveness of SHS in reducing frictional drag can be further understood through its ability to sustain a Cassie state\cite{cassie1944wettability}. In this state, a fluid rests on top of the surface structures without fully penetrating the microscopic cavities, forming a stable air layer between the fluid and the solid surface. This air layer significantly increases the effective slip length, thereby reducing shear stress at the interface. The stability and efficacy of the Cassie state depend strongly on the geometry and spacing of the microstructures, as well as on operating conditions such as flow velocity and pressure.

The actual structure of a superhydrophobic surface varies depending on its fabrication method. Techniques such as laser processing\cite{dunn2016laser}, lithography\cite{berendsen2009superhydrophobic}, and chemical etching\cite{qian2005fabrication} result in different surface morphologies, including groove-type patterns, regularly arranged post-type patterns, and randomly distributed roughness features. Typically,  previous studies assumed idealized surface geometries based on the fabrication approach, with numerical simulations being conducted separately for each structure. Consequently, past research has largely focused on isolated configurations, and a unified analysis of structural parameters across different SHS geometries remains lacking.

Numerous numerical studies have explored various SHS structure designs, including ridges\cite{park2013numerical, turk2014turbulent, jelly2014turbulence, im2017comparison}, transverse ridges\cite{martell2010analysis}, structured posts\cite{seo2016scaling, nguyen2024numerical}, and randomized posts\cite{seo2018effect}. However, most existing studies have primarily focused on a specific type of structure or idealized conditions, lacking comprehensive comparisons and systematic parameter analysis. Previous studies have categorized SHS patterns primarily into ridge-type, transverse ridge-type, and post-type structures with the spatial arrangement of slip (air-water interface) and no-slip (solid) boundary conditions based on the pattern chosen. However, these studies have generally treated each pattern type separately, often neglecting to elucidate any systematic relationship between the structural parameters that define the different types of SHS. As a result, a unified understanding of how structural spacing parameters affect drag reduction remains limited.

In this study, we propose a unified approach to investigate the drag reduction effects across various SHS patterns by parameterizing the spacing between posts in both streamwise ($d_1$) and spanwise ($d_2$) directions. By systematically varying these parameters within a range defined relative to post thickness ($0 \leq d_i \leq 2w$, where $i=1, 2$), we aim to explore a continuous spectrum of SHS structures encompassing the characteristics of ridge, transverse ridge, and post-type patterns.

To robustly quantify the uncertainties associated with these structural spacing parameters, we adopted the Latin hypercube sampling (LHS) method coupled with polynomial chaos expansion (PCE)\cite{xiu2007efficient,xiu2005high}. Direct numerical simulation (DNS) based on the spectral element method, implemented in NekRS, was employed to accurately capture the turbulence dynamics and precisely quantify drag reduction effects.

This unified and systematic approach not only facilitates the prediction of drag reduction performance across diverse SHS configurations but also identifies which geometric parameters exert the most significant influence on drag reduction efficacy. The outcomes of this study are expected to provide comprehensive insights and practical guidelines for designing optimized SHSs with predictable drag reduction capabilities under uncertain operational conditions.

\section{Methodology}
\subsection{\label{sec:Governing_eqn}Numerical Methods}
\begin{figure*}[ht!]
\includegraphics[width=1.0\linewidth]{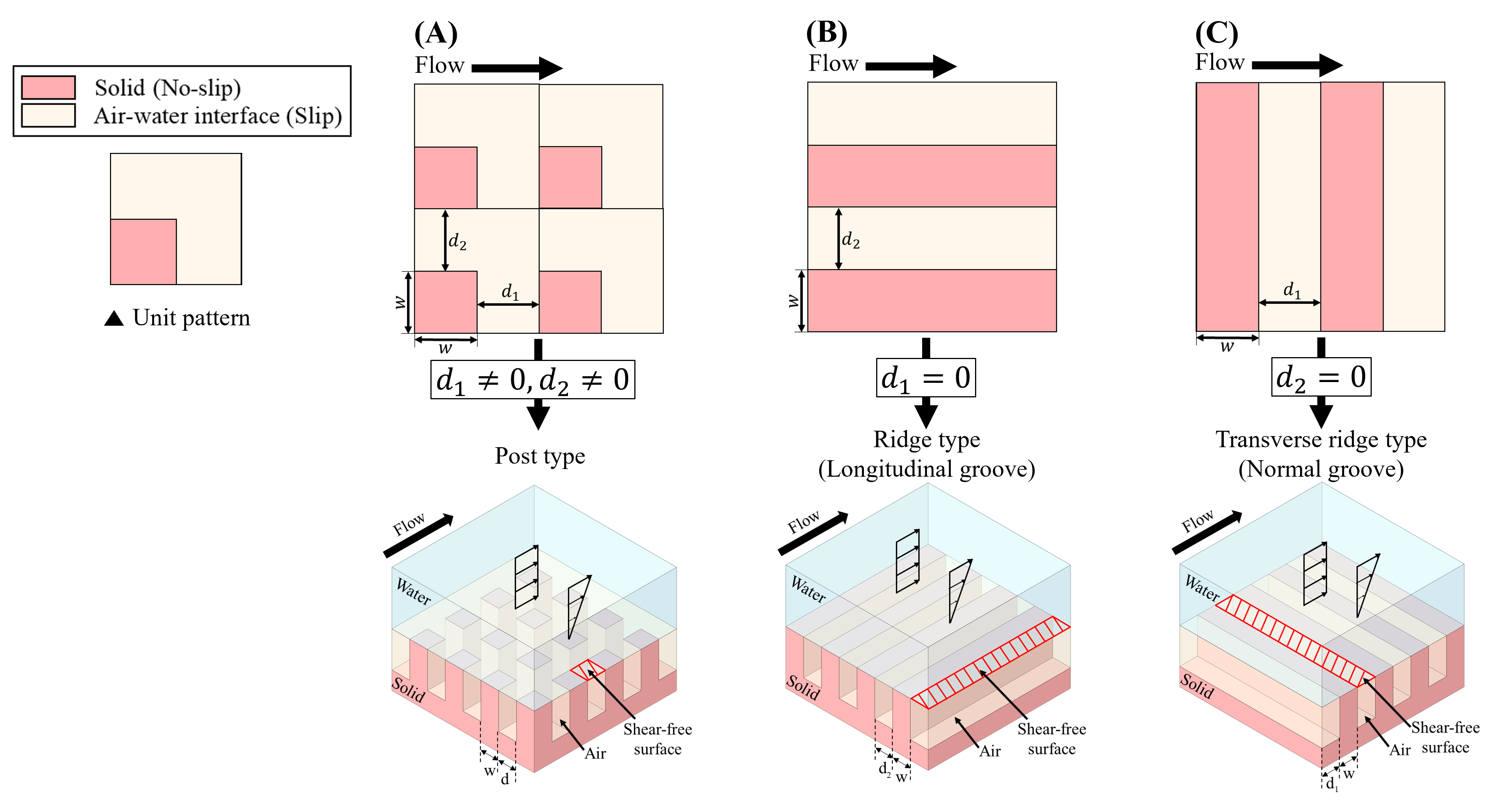}
\caption{\label{fig:SHS_structure}Schematic diagram of the various superhydrophobic structure patterns created by the parameterized structure spacing ($d_1$ : streamwise spacing, $d_2$ : spanwise spacing) of each unit pattern.}
\end{figure*}
The high-fidelity turbulent flow simulations presented in this study were conducted using NekRS, open-source GPU-oriented spectral element CFD code\cite{deville2002high} developed at Argonne National Laboratory by Fischer et al.\cite{fischer2022nekrs}. NekRS builds upon the well-established open-source code Nek5000\cite{fischer2008nek5000}, inheriting its established pre- and post-processing capabilities, such as mesh generation and data visualization. The simulations performed in this work use the incompressible, constant-properties Navier–Stokes equations in dimensionless form:

\begin{equation}\label{eqn:ge1}
\nabla\cdot{\textbf{u}}^*=0
\end{equation}
\begin{equation}\label{eqn:ge2}
\frac{\partial \textbf{u}^*}{\partial t^*} + \textbf{u}^* \cdot \nabla \textbf{u}^* = -\nabla p^*+ \frac{1}{Re} \Delta\textbf{u}^*  +\textbf{f}^*_{PG}
\end{equation}
\\

Equation (\ref{eqn:ge1}) is a non-dimensional continuity equation while equation (\ref{eqn:ge2}) is a non-dimensional Navier-Stokes equation. All the parameters are normalized by using the channel half height $h$ as the reference length and the bulk velocity $U_b$ as the reference velocity. The normalized parameters are expressed with a superscript *. The variables $u^*$ and $p^*$ are the velocity vector and pressure, respectively. The term $f^*_{PG}$ denotes the constant pressure gradient imposed along the channel, which was specifically adjusted for each of the friction Reynolds numbers ($Re_\tau$) considered in this study: 180, 395, and 590.

\subsection{\label{sec:Settings}Computational settings}
The computational domain considered in this study is a horizontal channel configuration. Consistent with the study by Martell et al.\cite{martell2009direct,martell2010analysis}, the bottom wall was set as an SHS, while the top wall was given as a no-slip boundary condition. Periodic boundary conditions were applied in both the streamwise and spanwise directions. The constant pressure gradient was adjusted according to each target friction Reynolds number($Re_\tau$).
The SHS on the bottom wall was modeled as a post-type pattern with alternating no-slip and no-shear boundary conditions. The thickness of the posts ($w$) was fixed at 0.1875$h$, and the spacing between posts was denoted as $d_1$ in the streamwise direction and $d_2$ in the spanwise direction. 

Figure \ref{fig:SHS_structure} presents a schematic diagram showing the various SHS configurations considered in this study. The patterns are composed of alternating no-slip (solid) and slip (air–water interface) regions arranged in repeating unit-cell structures. The slip regions are assumed to be a flat surface. The geometric configuration is defined by two parameters: the streamwise spacing $d_1$ and the spanwise spacing $d_2$. 
Depending on the values of $d_1$ and $d_2$, three representative SHS types can be systematically represented : post-type SHS, ridge-type SHS, and transverse ridge-type SHS. Figure \ref{fig:SHS_structure}(A) shows the post-type SHS pattern, in which both $d_1\ne0$ and $d_2\ne0$, resulting in discrete posts surrounded by slip regions. Figure \ref{fig:SHS_structure}(B) illustrates the ridge-type (longitudinal groove) SHS configuration, where the streamwise spacing is zero ($d_1=0$), forming continuous ridges aligned with the flow direction. Figure \ref{fig:SHS_structure}(C) corresponds to the transverse-ridge-type (normal groove) SHS pattern, where the spanwise spacing is zero ($d_2=0$), producing ridges perpendicular to the flow direction.
The 3D views further highlight the underlying air pockets that form beneath the air-water interface to produce shear-free surfaces, which play a critical role in reducing wall shear stress in turbulent flows. This unified parametrization enables a continuous transition among the various SHS geometries and provides a systematic framework for investigating their influence on drag reduction.

Table. \ref{tab:table1} shows the computation domain and grid information. In the computational domain, the unit SHS patterns are periodically repeated.
In this study, 30 different SHS geometries were investigated by changing the post spacing parameters($d_1, d_2$) for a specific friction Reynolds number. 
The numbers of unit patterns ($N_{unit, x}\times N_{unit, z}$) are 16 and 8 in the streamwise direction and spanwise direction, respectively. Therefore, the computational domain size ($L_x\times L_y\times L_z$) is  $16L_{unit, x}\times 2h \times 8L_{unit, z}$ for each case. The total computational domain varied depending on the SHS structure spacing: $d_1, d_2$.  

The simulation domain was divided into hexahedral elements and discretized into \textit{32$\times$24$\times$16} macro mesh elements. Each element is divided by the polynomial order(\textit{$N$}) into Gauss-Lobatto-Legendre(GLL) grid points. The polynomial orders were determined as 7th order and 9th order. 
Therefore, the total number of elements for each is 4.2 million and 9.0 million, respectively. The elements are uniformly distributed in the streamwise and spanwise directions. Rezaeiravesh et al.\cite{rezaeiravesh2021numerical} defined $\Delta x^+$ and $\Delta z^+$ as the average inner-scaled distances between the GLL points in the respective streamwise and spanwise directions, as follows : 
\begin{equation}\label{eqn:delta_x+}
\Delta x^+=\frac{L_x/h}{N_xN}Re_\tau
\end{equation}
\begin{equation}\label{eqn:delta_z+}
\Delta z^+=\frac{L_z/h}{N_zN}Re_\tau
\end{equation}
The average grid spacings in the streamwise direction $\Delta x^+$ and spanwise grid spacing $\Delta z^+$, are 4.8 wall units for $Re_\tau=180$, 8.2 for $Re_\tau=395$, and 12.3 for  $Re_\tau=590$. To resolve the smallest vortices near the wall, the grid is refined in the wall-normal direction, with spacing increasing gradually away from the wall. The grid spacing near the wall in the normal direction $\Delta y_w^+$ is about 0.089 and while the largest grid spacing in the same direction $\Delta y_c^+$ is 7.65 near the center of the channel. This grid resolution is comparable to those found in previous literature for turbulent channel DNS using Nek5000\cite{kim2025euler}.
The Courant–Friedrichs–Lewy (CFL) number was set below 0.5, while the time step size varied depending on the CFL number. 
Initially, the flow was driven by a streamwise constant pressure gradient corresponding to the desired Reynolds number. In a fully developed channel, the average wall shear stress $\tau_w$ is directly related to the average channel pressure gradient, $\partial P / \partial x$ by $\tau_w = (2/L_y )(\partial P /\partial x)$\cite{pope2000turbulent}. The frictional Reynolds number $Re_\tau$ was set at 180 as calculated by $Re_\tau=\rho u_\tau h/\mu$. 

The $P_N-P_N$ formulation is used for velocity and pressure with polynomial order $N$. An operator-integration-factor splitting (OIFS) scheme was adopted for temporal discretization. The linear terms were solved implicitly with 3rd-order backward differentiation(BDF3), while non-linear terms were solved using a characteristics-based explicit scheme. The OIFS method allows for larger time steps than would be allowed with regular explicit/extrapolation schemes \cite{fischer2003implementation}.

\begin{table}
\caption{\label{tab:table1}Computational domain and grid information.}
\begin{ruledtabular}
\begin{tabular}{l|ccc}
\multirow{2}{*}{Parameter name}&\multicolumn{3}{c}{$Re_\tau$}\\
\cline{2-4}
& $180$ & $395$ & $590$\\
\hline
Channel configuration\footnote{$L_x=N_{unit,x}L_{unit, x},  L_y=2h,  L_z=N_{unit, z}L_{unit, z}$}&\multicolumn{3}{c}{$L_x \times L_y \times L_z$}\\ 
Unit pattern size, $L_{unit, x} \times L_{unit, z}$ &\multicolumn{3}{c}{$(w+d_1) \times (w+d_2)$}\\
Unit patterns, $N_{unit,x} \times N_{unit, z}$ &\multicolumn{3}{c}{$16 \times 8$}\\
Spectral elements, $N_x \times N_y \times N_z$ & \multicolumn{3}{c}{$32\times24\times16$} \\ 
Polynomial order, $N$ & 7 & 9 & 9 \\
The number of cells, $EN^3$ & $4.2\times 10^6$ & $9.0\times 10^6$ & $9.0\times 10^6$ \\
Wall unit near wall, $\Delta y^+_{wall}$ & 0.089 & 0.194 & 0.292
\end{tabular}
\end{ruledtabular}
\end{table}

\subsection{\label{sec:UQ_Theory}Uncertainty quantification}
\begin{figure*}
\includegraphics[width=1.0\linewidth]{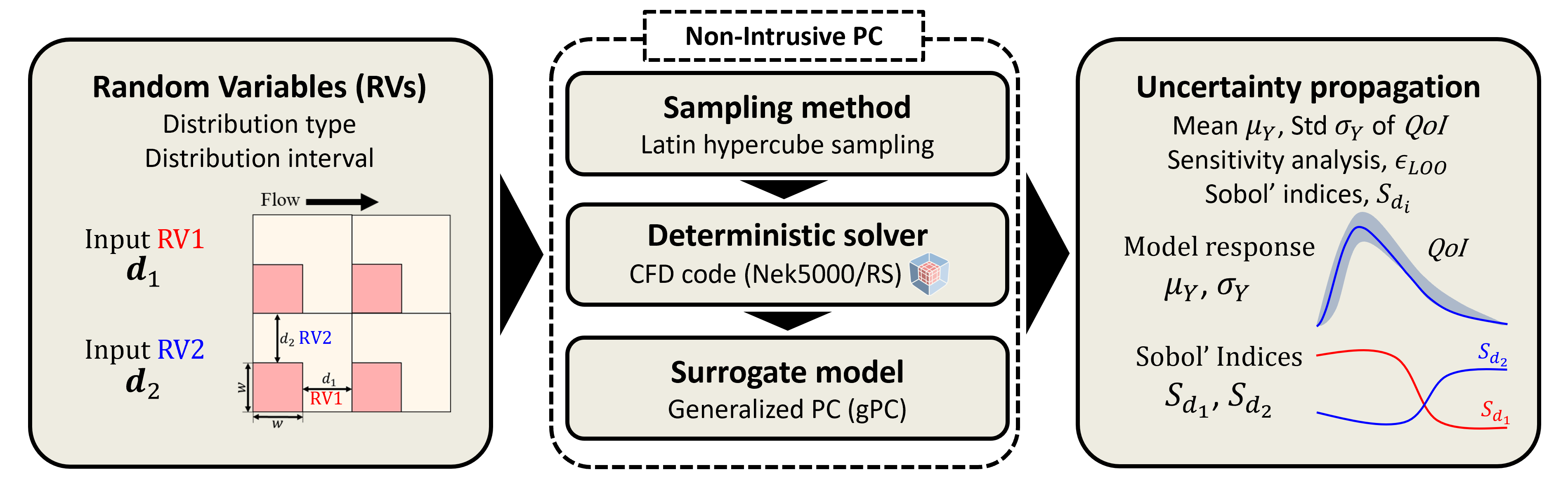}
\caption{\label{fig:UQ_procedure}The procedure of uncertainty quantification for two input random variables ($d_1, d_2$) of superhydrophobic structure}
\end{figure*}

 There are two approaches for Uncertainty quantification(UQ) method; the forward problem and the inverse problem. The forward problem was conducted in this work, and a schematic diagram of this process is shown in Fig.\ref{fig:UQ_procedure}. There are three steps in the forward problem. First of all, the probability distribution function and interval of the input random variables are assumed. In the second step, the non-intrusive polynomial chaos method is conducted with a deterministic CFD solver; in our case NekRS. Depending on the input random variables, the uncertainty propagation is then calculated during the last step with the mean and standard deviation. In Sect. \ref{sec:PCNIPC}, the point-collocation non-intrusive polynomial chaos(PC-NIPC) method will be explained in detail. In Sect. \ref{sect:Loo error}, the sensitivity analysis for the surrogate model built by the polynomial chaos method will also be presented.

\subsubsection{\label{sec:PCNIPC}Point-Collocation Non-Intrusive Polynomial Chaos}
The Point-collocation non-intrusive polynomial chaos(PC-NIPC) approach, proposed by Hosder et al.\cite{hosder2007efficient}, was adopted in this work. The advantage of PC-NIPC is that it doesn’t require the modification of the governing equations. The less number of samples needed than the numerical quadrature NIPC in increasing the number of random variables. The equation for the number of samples is written as :
\newline
\begin{equation}\label{eqn:LHS}
P+1=n_p \frac{(n+p)!}{n!p!}
\end{equation}
\newline
Here, the oversampling rate $n_p$ was chosen as $2$ for PCE accuracy\cite{hosder2007efficient}. The polynomial order and the number of random variables are represented by $p$ and $n$, respectively. The samples are selected by using the Latin hypercube sampling(LHS) method\cite{helton2003latin}. These samples are used for the equation below :
\newline
\begin{equation}\label{eqn:PCE}
    Y(X,\xi_j)=\sum_{n=0}^{\infty}\alpha_n(X)\Psi_n(\xi_j),\quad j=0,1,2,\cdots,P
\end{equation}
\newline
Here, $\alpha_n$ is the vector of polynomial chaos coefficients and $\Psi_n$ is an element of an orthogonal family. The response corresponding to the random variable is written as $Y$. $\xi_j$ represents a vector of random variables. In equation (\ref{eqn:PCE}), $Y$ and $\Psi_n$ are the given values. By applying the Gram-Schmidt method,  $\alpha_n$ can be obtained as: 
\newline
\begin{equation}\label{eqn:GS_matrix}
\left[
\begin{array}{c}
 Y_0\\
 Y_1\\
 \vdots \\
 Y_{P-1}\\
 Y_P
\end{array}
\right] = 
\left[
\begin{array}{cccc}
 \Psi_0(\xi_0)&\Psi_1(\xi_0)&\cdots&\Psi_P(\xi_0)\\
 \Psi_0(\xi_0)&\Psi_1(\xi_1)&\cdots&\Psi_P(\xi_1)\\
 \vdots&\vdots&\ddots&\vdots\\
 \Psi_0(\xi_{P-1})&\Psi_1(\xi_{P-1})&\cdots&\Psi_P(\xi_{P-1})\\
 \Psi_0(\xi_{P})&\Psi_1(\xi_{P})&\cdots&\Psi_P(\xi_P)\\
\end{array}
\right]
\left[
\begin{array}{c}
\alpha_0\\
\alpha_1\\
\vdots\\
\alpha_{P-1}\\
\alpha_P\\
\end{array}
\right]
\end{equation}
\newline
The element of an orthogonal family $\Psi_n$ is decided by the type of the input random variable distribution. Table \ref{table:PCE_basis_fn} shows the basis function of the gPCE\cite{Xiu+2010}. In the present work, the Hermite basis function is adopted due to its Gaussian distribution.

\begin{table}
\centering
\caption{gPCE basis function family}
\begin{ruledtabular}
\begin{tabular}{cccc}
 \textbf{Distribution} & \textbf{gPCE basis} & \textbf{Interval} & \textbf{Weight function}  \\
 \hline
 Gaussian & Hermite & ($-\infty$, $\infty$) & $exp(-\frac{x^2}{2})$  \\
 Uniform & Legendre & [$-1$, $1$] & $1$ \\
 Gamma & Laguerre & [$0$, $\infty$] & $exp(-x)$  \\
\end{tabular}\label{table:PCE_basis_fn}
\end{ruledtabular}
\end{table}

\subsubsection{\label{sect:Loo error}Leave-One-Out Cross-Validation Error}
The leave-one-out (LOO) cross-validation error, $\epsilon_{LOO}$ is used to estimate the error and compare the accuracy of the PCE. The LOO cross-validation error is designed to overcome the over-fitting limitation of the normalized empirical error using cross-validation. 
\begin{equation}
    \epsilon_{LOO}=\frac{\sum_{i=1}^{N}{((Y(x^{(i)})-Y^{PC}(x^{(i)})/(1-h_i))}^2}{\sum_{i=1}^N {(Y(x^{(i)})-\hat{\mu}_Y)}^2}
\end{equation}
where $\textbf{h}_i$ is the $i$th component of the vector, which is expressed as:
\newline
\begin{equation}
    \textbf{h}=diag(\textbf{A}(\textbf{A}^{\top}\textbf{A})^{-1}\textbf{A}^{\top})
\end{equation}
\begin{equation}
    \textbf{A}_{ij}=\Psi_j(\xi_i)
\end{equation}
where $i=1,\cdots, n$, and $j=0, \cdots, P-1$.

\subsubsection{\label{sec:UQ_setting}Uncertainty quantification settings}
In this work, two random input variables were selected: the streamwise distance($d_1$) and spanwise distance($d_2$) between the post micro-structures. The probability distributions were assumed to be Gaussian distributions. The mean values ($\mu$) are for the SHS structure width($w$), and the standard deviations ($\sigma$) were set as 0.33$\mu$ to prevent these values from being negative. The surrogate models are built with the polynomial chaos expansion(PCE) method. The surrogate models are of the 2nd, 3rd, and 4th polynomial orders. Therefore, the number of samples are calculated as 12, 20, 30 by Equation (\ref{eqn:LHS}). The entire UQ analysis was carried out using UQLab\cite{marelli2014uqlab}.

\section{Deterministic Results}
\subsection{\label{sec:DM_Validation}Validation}
\begin{figure*}
    \centering
    \includegraphics[width=1\linewidth]{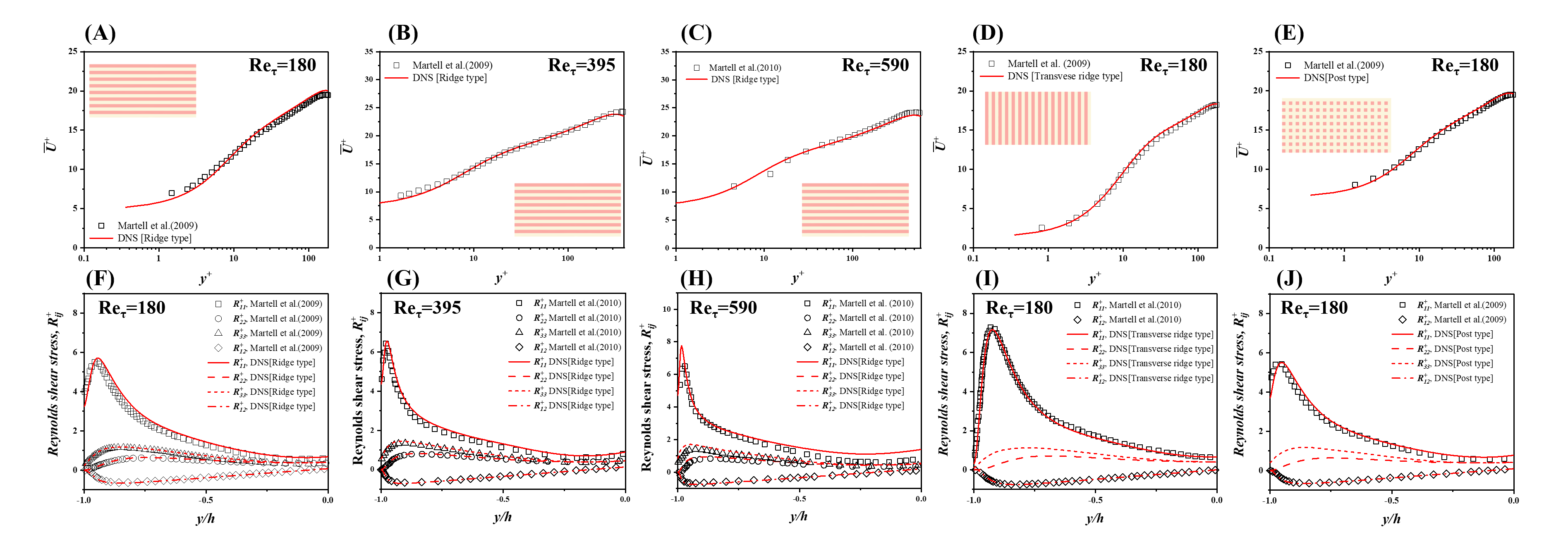}
    \caption{Validation of DNS results according to the SHS type and shear Reynolds number. (A-E) velocity profile, (F-J) Reynolds stress profile.}
    \label{fig:validation}
\end{figure*}

The mean streamwise velocity and Reynolds shear stress profiles for various SHS configurations are compared against Martell et al.\cite{martell2009direct,martell2010analysis}. Figure \ref{fig:validation} (A-C) shows the streamwise velocity profiles for ridge-type SHS at shear Reynolds numbers $Re_\tau=180$, $395$, and $590$, respectively. Figure \ref{fig:validation} (D, E) present streamwise velocity profile for transverse-ridge type and post-type SHS at $Re_\tau=180$, respectively. Corresponding Reynolds shear stress components $R^+_{11}, R^+_{22}, R^+_{33},$ and $R^+_{12}$, are shown in (F-J) demonstrating good agreement with the reference results for each configuration. These results confirm that the computational settings are sufficient to accurately capture the physical phenomena of interest. 


\subsection{\label{sec:DM_drag_redution}Drag reduction}
\begin{figure}
    \centering
    \includegraphics[width=1\linewidth]{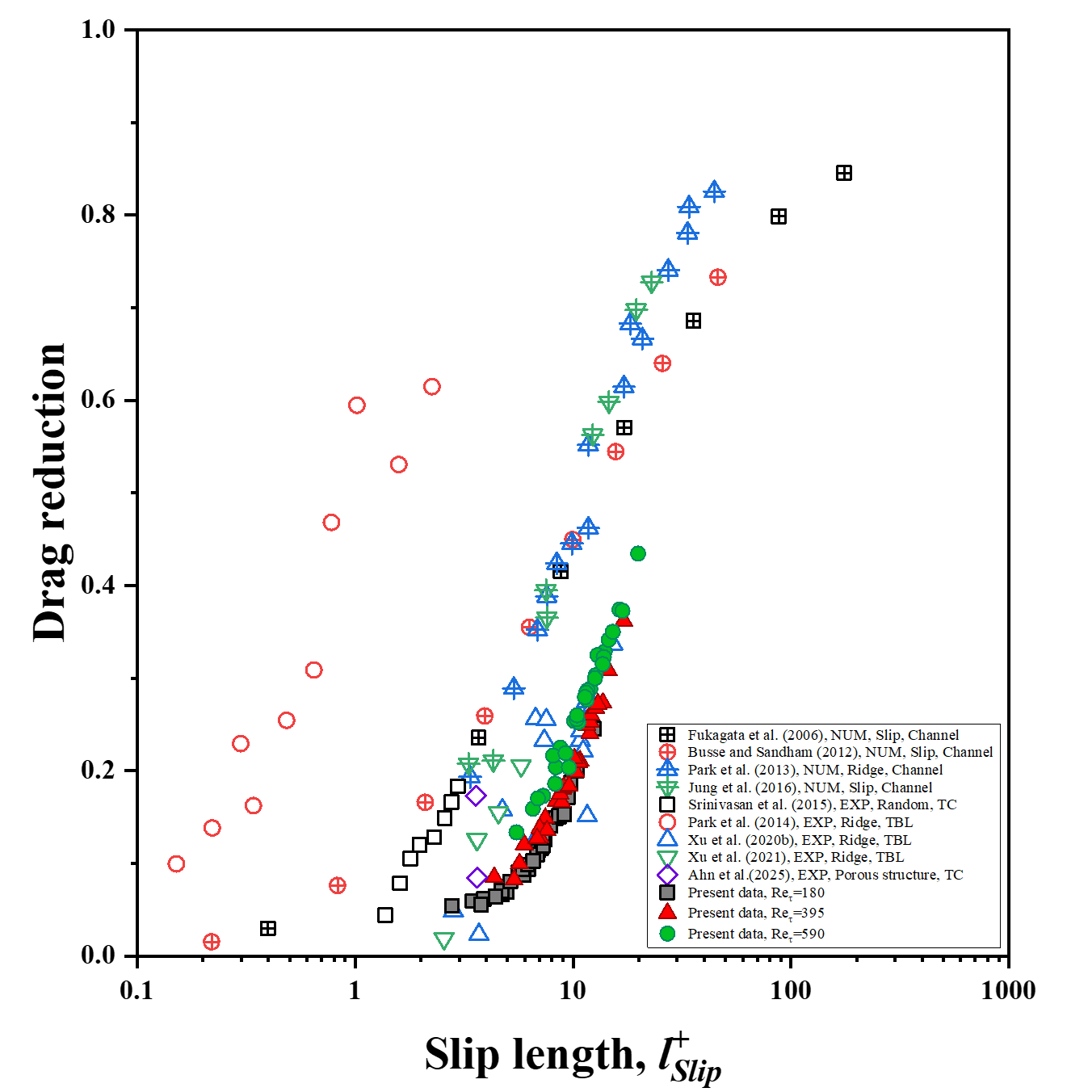}
    \caption{Graph of drag reduction according to the dimensionless slip length($l^+_{Slip}$). The present results are plotted with black rectangles ($Re_\tau=180$), red triangles ($Re_\tau=395$), and green circles ($Re_\tau=590$). This figure is regenerated from Park et al.\cite{park2021superhydrophobic}} 
    \label{fig:DR_SL}
\end{figure}
Figure \ref{fig:DR_SL} shows the relationship between the drag reduction and dimensionless slip length($l^+_{Slip}$), comparing our results with previous numerical and experimental studies. These previous numerical studies were conducted by applying an effective slip length imposed condition\cite{fukagata2006theoretical,busse2012influence,jung2016effects} , and ridge-type SHS\cite{park2013numerical}. By the experiments, the ridge-type SHS\cite{park2014superhydrophobic,xu2020superhydrophobic,xu2021superhydrophobic} were investigated, and the random distribution SHS\cite{srinivasan2015sustainable,ahn2025characteristics} were investigated.
A general trend can be observed across all datasets: drag reduction increases monotonically with slip length, regardless of Reynolds number or surface pattern type. The present simulation results at $Re_\tau=180,$ $395$, and $590$ (represented by black squares, red triangles, and green circles, respectively) are comparable to previous studies, and are consistently shifted toward higher slip lengths. This trend is consistent with the findings of Martell et al.\cite{martell2009direct} in which post-type SHS were shown to yield greater slip velocity than the ridge-type SHS with the same microstructure distances. According to Martell et al.\cite{martell2009direct}, post-type SHS produced about 20$\%$ higher slip velocity than ridge-type SHS.


\subsection{\label{sec:DM_domain_test}Anisotropy invariant map}
To investigate the effect of SHS patterns on near-wall turbulence structures, an anisotropy invariant map(AIM) was adopted for analysis of the Reynolds stress anisotropy. The AIM provides a compact and physically interpretable representation of the turbulence anisotropy by projecting the second-order moments of velocity fluctuations into an invariant space.
The anisotropy tensor $b_{ij}$ is defined as the deviatoric part of the normalized Reynolds stress tensor:

\begin{equation}
    b_{i j}=\frac{\overline{u_{i}^{\prime} u_{j}^{\prime}}}{2 k}-\frac{1}{3} \delta_{i j},
\end{equation}
where $k=\frac{1}{2} \overline{u_{i}^{\prime} u_{i}^{\prime}}$ is the turbulent kinetic energy, and $\delta_{ij}$ is the Kronecker delta. The invariants of the anisotropy tensor, $\mathrm{II}$ and $\mathrm{III}$, are then computed as:
\begin{equation}
    \mathrm{II}=b_{i j} b_{j i}, \quad \mathrm{III}=b_{i j} b_{j k} b_{k i}
\end{equation}

These two invariants define a two-dimensional invariant space bounded by a triangle, commonly referred to as the Lumley triangle\cite{lumley1977return}, which encompasses all physically realizable states of turbulence. However, due to the overlap of data points that occurs when visualizing two-dimensional distributions, the Lumley triangle becomes inadequate for clearly representing anisotropy variations. To address this limitation, the barycentric map proposed by Emory and Iaccarino\cite{emory2014visualizing} is employed in this study.
The barycentric map is defined in terms of the eigenvalues $\lambda_i$ of the anisotropy tensor $b_{ij}$\cite{pope2000turbulent}, with the following coordinates:

    \begin{equation}
    C_{1c}=\lambda_1-\lambda_2
    \end{equation}
    \begin{equation}
    C_{2c}=2(\lambda_2-\lambda_3)
    \end{equation}
    \begin{equation}
    C_{3c}=3\lambda_3+1
    \end{equation}
    \begin{equation}
    x_B=C_{1c}x_{1c}+C_{2c}x_{2c}+C_{3c}x_{3c}=C_{1c}+\frac{1}{2}C_{3c}
    \end{equation}
    \begin{equation}
    y_B=C_{1c}y_{1c}+C_{2c}y_{2c}+C_{3c}y_{3c}=\frac{\sqrt{3}}{2}C_{3c}
    \end{equation}
Here, the vertices of the barycentric triangle correspond to various limiting states of turbulence: one-component ($x_{1c}=(1,0)$), two-component ($x_{2c}=(0,0)$), and isotropic turbulence ($x_{3c}=(1/2,\sqrt3/2)$). Each turbulence state is represented as a convex combination of these vertices using the barycentric coordinates ($x_B,y_B$).

By mapping SHS-induced turbulent flows onto the AIM, we are able to identify how variations in surface patterning (e.g., streamwise and spanwise spacing) modify the anisotropy characteristics of near-wall turbulence, which in turn provides deeper insight into how surface geometry influences the local turbulence structure and the underlying mechanisms of drag reduction.

\begin{figure*}
    \centering
    \includegraphics[width=1.0\linewidth]{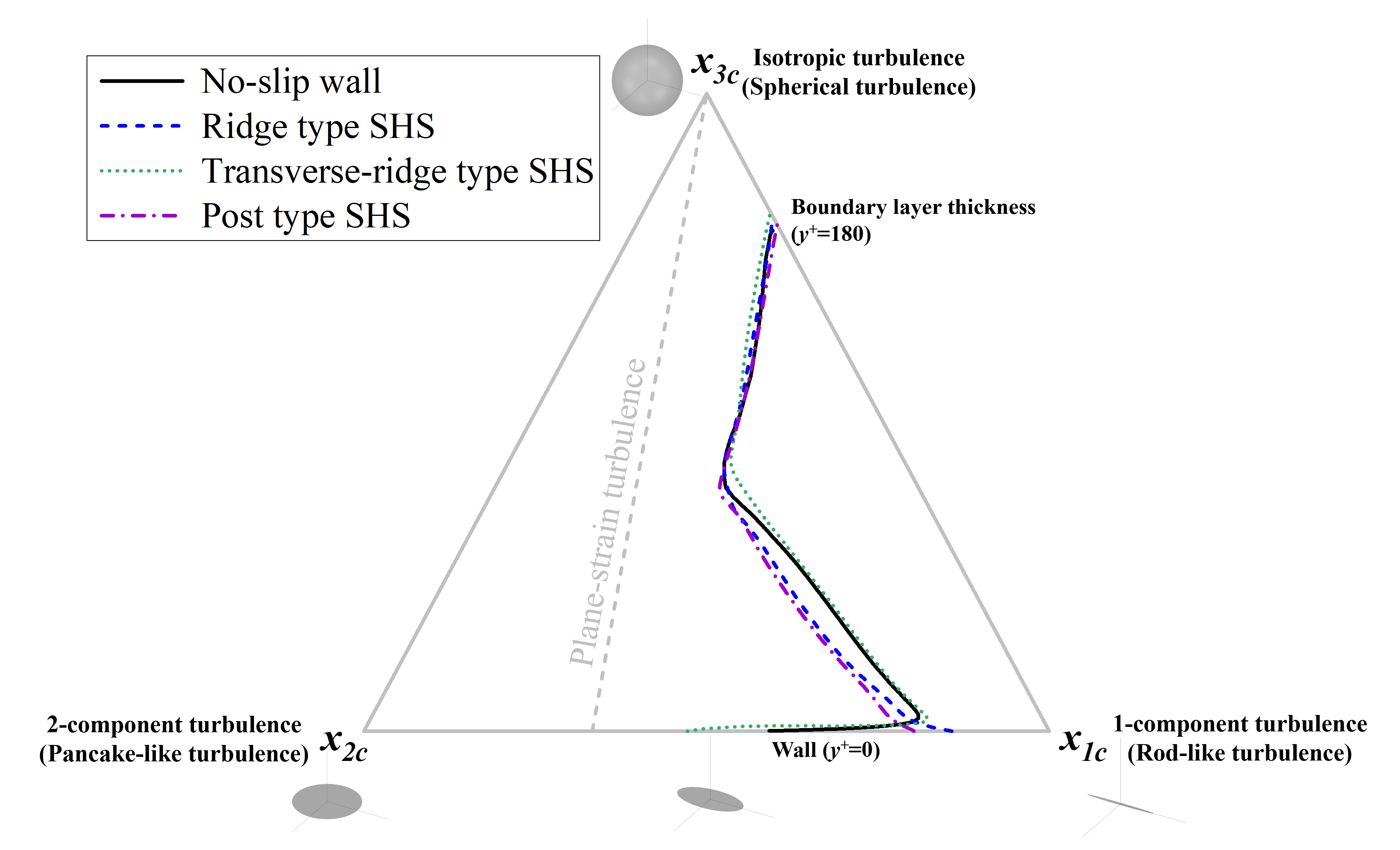}
    \caption{The barycentric anisotropy invariant map for representative cases(\protect\blackline~No-slip wall channel, \protect\bluedash~Ridge type SHS channel, \protect\greendot~Transverse ridge type SHS channel, \protect\purpledashdot~Post type SHS channel) at $Re_\tau=180$.}
    \label{fig:AIM_Re180}
\end{figure*}

Figure \ref{fig:AIM_Re180} presents the anisotropy invariant map (AIM) for various wall boundary conditions at $Re_\tau=180$, including no-slip walls and three representative SHS configurations: ridge type, transverse-ridge type, and post type. The AIM trajectory describes the evolution of the turbulence anisotropy across the wall-normal direction, away from the wall ($y^+=0$) to the boundary layer thickness ($y^+\simeq 180$).
In the no-slip wall case, the trajectory begins at the lower edge of the triangle, close to the two-component turbulence limit, which reflects the highly anisotropic state near the wall where wall-normal fluctuations are strongly suppressed. As the wall-normal distance increases from the wall, the trajectory in the AIM initially moves toward the lower-right corner, representing a one-component turbulence state, where streamwise velocity fluctuations dominate and wall-normal and spanwise components are strongly suppressed. This behavior reflects the presence of elongated streamwise streaks and highly anisotropic structures in the viscous sublayer. As the distance increases further through the buffer and logarithmic layers, the trajectory bends away from the one-component limit and gradually ascends toward the isotropic vertex. This transition indicates a progressive redistribution of turbulent kinetic energy among all three components, with turbulence becoming increasingly isotropic in the outer region. The overall shape of the trajectory closely aligns with those reported for canonical turbulent channel flows\cite{hoyas2006scaling, emory2014visualizing}. This characteristic shape captures the classical transition from wall-dominated anisotropy to more isotropic behavior in the outer layer.

All SHS cases initiate their trajectories from the lower edge of the triangle, similar to the no-slip case. However, the AIM trajectories for SHS cases begin to notably diverge from the no-slip case, especially in the near-SHS region. For both the ridge-type and post-type SHS, the trajectories do not exhibit a transitional path from two-component to one-component turbulence. Instead, they originate directly from a strong one-component turbulence state at $y^+=0$, and gradually move toward the isotropic turbulence as the wall-normal distance increases. This behavior is consistent with the observations of Watanabe et al.\cite{watanabe2017drag} and Elboth et al.\cite{elboth2012flow}, both of whom investigated ridge-type SHS. In contrast, the transverse ridge-type SHS exhibits a distinctly different behavior, starting with an even stronger two-component turbulence state at the SHS compared to the no-slip case. This observation underscores the significant impact of surface patterning on the turbulence structure in the vicinity of a SHS.

The AIM analysis provides insight into how SHS-induced modifications in turbulence anisotropy relate to changes in near-wall coherent structures. In the no-slip wall case, the typical trajectory near the two-component turbulence limit is associated with the presence of highly organized streamwise streaks and quasi-streamwise vortices, which are responsible for sustaining wall-bounded turbulence and generating significant frictional drag. In contrast, the altered trajectories observed for SHS cases indicate a breakdown or weakening of coherent structures. Furthermore, the transverse ridge-type SHS, which maintains stronger two-component turbulence near the wall, shows a more limited degree of drag reduction. This correlation reinforces the idea that the effectiveness of a SHS is strongly tied to its ability to attenuate coherent structure formation near the wall. The suppression of vortical structures over SHS has also been reported in previous studies\cite{min2004effects, martell2009direct, park2013numerical, nguyen2022numerical}.

\section{Uncertainty Quantification Results\label{sec:UQ_results}}

In Section~\ref{sec:UQ_results}, the results of the uncertainty quantification (UQ) analysis are presented. Two input random variables (RVs) are considered: the streamwise directional spacing ($d_1$) and the spanwise directional spacing ($d_2$). Both RVs are assumed to follow Gaussian distributions. The detailed settings for the RVs are summarized in Table~\ref{tab:UQ_setting}.
The mean spacing ($\mu$) for both variables is set to $0.1875h$, which corresponds to the post width. The standard deviation ($\sigma$) is chosen such that the sampled values remain non-negative; accordingly, $\sigma$ is defined as $\pm0.33\mu$.
Surrogate models are constructed using polynomial chaos expansion (PCE), while the required number of samples is determined from Equation~\ref{eqn:LHS}. We consider 2nd-, 3rd-, and 4th-order polynomial surrogate models, requiring 12, 20, and 30 samples, respectively.

\begin{table*}
\caption{\label{tab:UQ_setting}The probability distributions of input random variables for uncertainty quantification.}
\begin{ruledtabular}
\begin{tabular}{cccc}
 \textbf{Input random variables(RVs)} & \textbf{Probability distribution} & \textbf{Mean,} $\mu$ & 
 \textbf{Standard deviation,} $\sigma$ \\
\hline
Streamwise spacing, $d_1$ & Gaussian distribution & $0.1875h$ & $\pm0.33\mu$  \\
Spanwise spacing, $d_2$ & Gaussian distribution & $0.1875h$ & $\pm0.33\mu$  \\
\end{tabular}
\end{ruledtabular}
\end{table*}

\begin{table*}
    \centering
    \caption{The LOO error and probability distribution of drag reduction regarding the PCE order for each Reynolds number.}
    \begin{ruledtabular}
    \begin{tabular}{cccccc}

        \textbf{Reynolds number}, $Re_\tau$ & \textbf{PCE order}, $p$ & \textbf{LOO error}, $\epsilon_{LOO}$ & \textbf{Mean}, $\mu_Y$ & \textbf{Std}, $\sigma_Y$ \\
        \midrule
        \multirow{3}{*}{180} & 2 & 0.09 & 11.3 & 4.64 \\
                             & 3 & 0.03 & 11.4 & 4.33 \\
                             & 4 & 0.41 & 11.7 & 4.52 \\
        \midrule
        \multirow{3}{*}{395} & 2 & 0.14 & 19.6 & 6.62 \\
                             & 3 & 0.11 & 19.5 & 6.64 \\
                             & 4 & 0.06 & 18.9 & 7.03 \\
        \midrule
        \multirow{3}{*}{590} & 2 & 0.08 & 27.2 & 6.93 \\
                             & 3 & 0.50 & 27.0 & 6.79 \\
                             & 4 & 0.06 & 26.7 & 7.99 \\

    \end{tabular}
    \end{ruledtabular}
    \label{tab:QoI_drag_reduction}
\end{table*}

\begin{table*}
\centering
\caption{The Sobol' indices of the input random variables ($d_1, d_2$) by Reynolds number, $Re_\tau$}
\begin{ruledtabular}
\begin{tabular}{cccccc}
\textbf{Reynolds number, $Re_\tau$} & \textbf{PCE order, $p$} &
\textbf{Sobol' index of $d_1$,} $S_{d_1}$ & \textbf{Sobol' index of $d_2$,} $S_{d_2}$ & \textbf{Complex Sobol' index,} $S_{d_1, d_2}$ \\
\midrule
180 & 3 & 0.123 & 0.849 & 0.029 \\
395 & 4 & 0.131 & 0.799 & 0.070 \\
590 & 4 & 0.058 & 0.813 & 0.129 \\
\end{tabular}
\end{ruledtabular}
\label{tab:UQ_Sobol}
\end{table*}

\subsection{\label{sec:UQ_DR}Drag reduction}
To quantify the uncertainty that comes from input random variables, three surrogate models of different polynomial order ($p$) are built. By comparing LOO error ($\epsilon_{LOO}$) of each surrogate model, the surrogate model with the closest response is determined. The quantity of interest (QoI) is drag reduction, as shown in the Table. \ref{tab:QoI_drag_reduction}. This table displays a probability distribution based on the PCE order and LOO error, which enables the quantification of the model response to the solution. 

Table~\ref{tab:QoI_drag_reduction} summarizes the leave-one-out (LOO) cross-validation errors and the probability distribution of drag reduction for various polynomial chaos expansion (PCE) orders at three different Reynolds numbers. The LOO error, denoted as $\epsilon_{LOO}$, quantifies the accuracy of each surrogate model. At $Re_\tau=180$, the 3rd-order PCE model achieves the lowest LOO error (0.03), indicating superior surrogate accuracy compared to the 2nd- and 4th-order models. At $Re_\tau=395$ and $Re_\tau=590$, the 4th-order model achieves the lowest LOO error of 0.06 for both, suggesting that the optimal PCE order varies slightly with Reynolds number. Therefore, the UQ results for $Re_\tau=180$ are derived from the 3rd-order model, and the UQ results for $Re_\tau=395, 590$ are derived from the 4th-order model.

Table \ref{tab:UQ_Sobol} shows the Sobol' index of each input variables($S_{d_1}$, $S_{d_2}$) and the second-order Sobol index($S_{d_1, d_2}$) corresponding to the three Reynolds numbers. For each Reynolds number, the Sobol' indices were compared using the most suitable PCE model order. Across all Reynolds numbers, the spanwise spacing $d_2$ consistently emerged as the dominant input variable. At $Re_\tau=180$, $S_{d_2}$ was approximately seven times more influential than $S_{d_1}$. Similarly, $S_{d_2}$ was about six times more dominant at $Re_\tau=395$, and about fourteen times more dominant at $Re_\tau=590$. These results indicate that, compared to $d_1$, $d_2$ plays a more significant role in the drag reduction induced by the superhydrophobic surface structures. The behavior of the interaction Sobol' index ($S_{d_1, d_2}$) is also interesting. As the Reynolds number($Re_\tau$) increases, the contribution of $S_{d_1, d_2}$ also increases gradually. Notably, at $Re_\tau=590$, $S_{d_1, d_2}$ exceeds the individual contribution of $S_{d_1}$, indicating a stronger coupling effect between the two input variables at higher Reynolds numbers.

\begin{figure}
    \centering
    \includegraphics[width=1.0\linewidth]{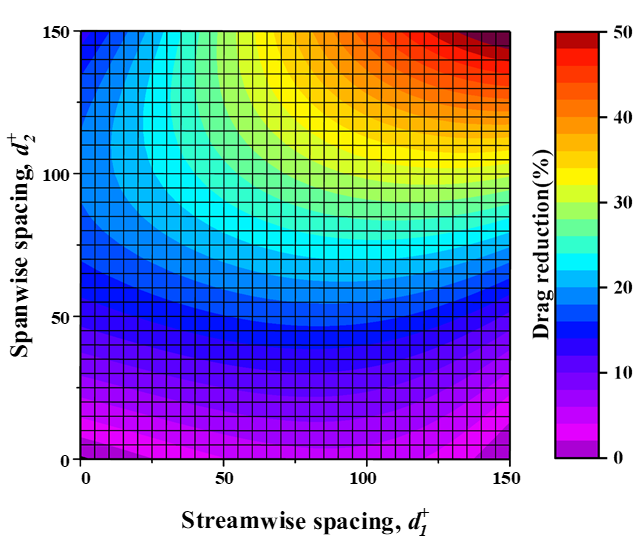}
    \caption{The response surface of drag reduction according to the two input RVs : $d_1^+, d_2^+$}
    \label{fig:Resonse_surface_DR}
\end{figure}

Figure \ref{fig:Resonse_surface_DR} presents the response surface of drag reduction as a function of two input variables. The streamwise ($d_1^+$) and spanwise ($d_2^+$) spacings of the SHS structures are expressed in wall units. For each Reynolds number ($Re_\tau=180$, $395$, and $590$), 1000 sample data points were generated from the surrogate model to construct the response surface. The response surface is represented using a polynomial surface, and the polynomial order was selected based on the coefficient of determination ($R^2$) and residual error. As the polynomial order increases from 1 to 6, the model accuracy improves, as indicated by the increasing values of $R^2$ and decreasing residual error. However, the improvement becomes marginal beyond the third-order model. For instance, the gain in adjusted $R^2$ from the third to sixth order is only about 0.018, despite a significant increase in model complexity. Thus, the third-order polynomial surface was selected for its balance between accuracy ($Adj. R^2 = 0.873$) and interpretability. The response surface reveals that variations in drag reduction are more sensitive to the spanwise spacing $d_2^+$ than to the streamwise spacing $d_1^+$. This indicates that $d_2^+$ is the more dominant parameter in determining drag reduction performance. Additionally, as both $d_1^+$ and $d_2^+$ increase, the solid fraction of the SHS decreases, leading to an increase in drag reduction.


\subsection{\label{sec:UQ_VP}Velocity profile}
\begin{figure*}
    \centering
    \includegraphics[width=\textwidth]{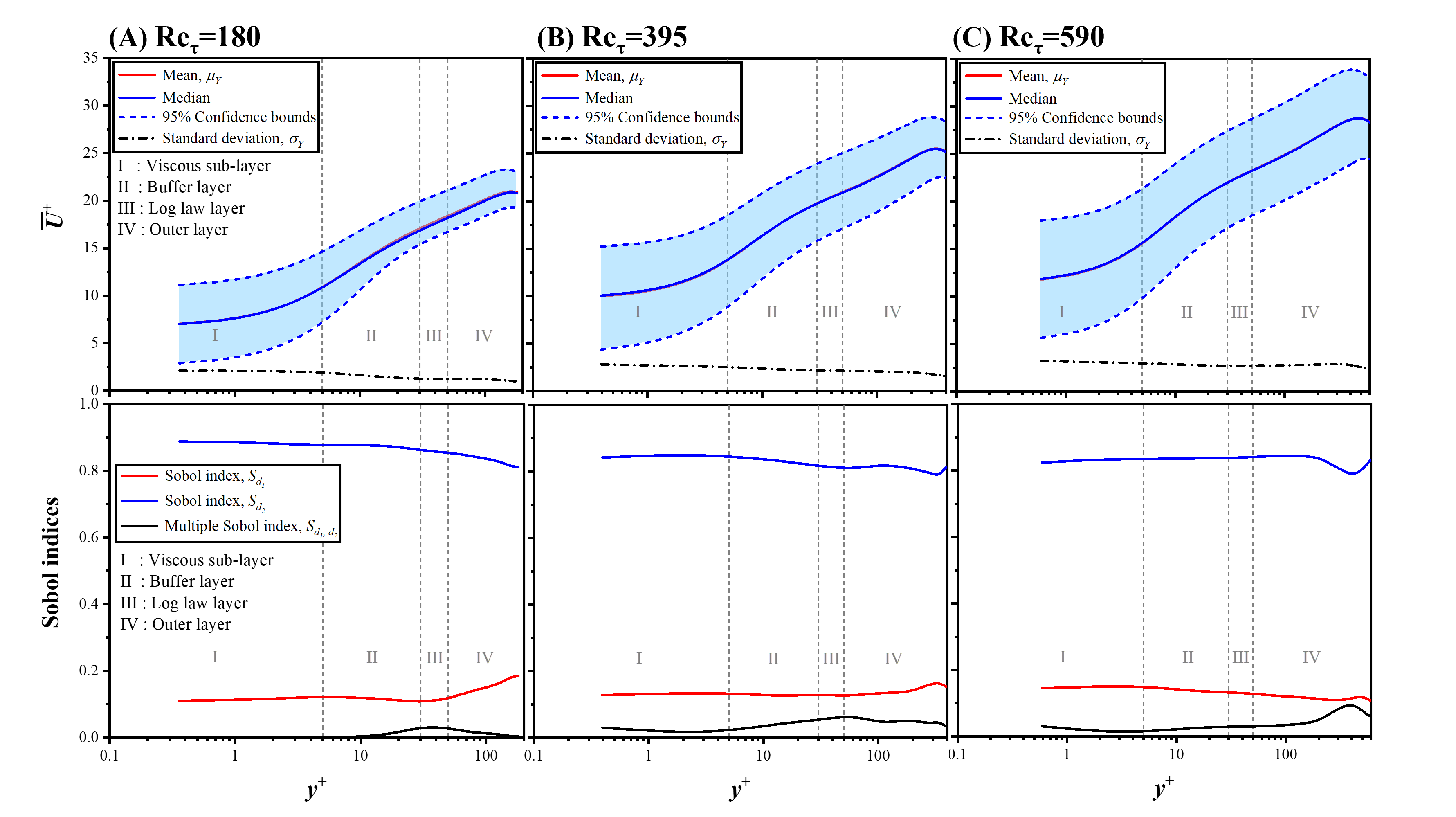}
    \caption{The uncertainty propagation of the velocity profile according to the shear Reynolds number($Re_\tau=180, 395, 590$)}
    \label{fig:UQ_Velocity_profile}
\end{figure*}

Figure \ref{fig:UQ_Velocity_profile} shows the results of the uncertainty propagation and sensitivity analysis of the streamwise velocity profile ($\bar{U}^+$) for three friction Reynolds numbers, $Re_\tau=180, 395$, and $590$. Vertical dashed lines indicate the characteristic sublayers of the turbulent boundary layer: I. viscous sublayer, II. buffer layer, III. log-law layer, and IV. outer layer.
In the upper panels, the shaded region represents the 95$\%$ confidence interval, bounded by blue dashed lines, while the black dashed-dotted line indicates the standard deviation profile. In all cases, the uncertainty introduced by variations in the geometric parameters ($d_1, d_2$) results in noticeable deviations in the velocity profile, particularly within the viscous sublayer region. The first thing that can be seen is that the uncertainty propagation has the largest interval near the SHS. This trend is consistent across all Reynolds numbers, though the absolute magnitude of uncertainty grows slightly with increasing $Re_\tau$. 
The lower panels present the Sobol' indices for the input variables $d_1$ (red), $d_2$ (blue), and their interaction term $S_{d_1,d_2}$ (black). For each Reynolds number, $d_2$ (spanwise spacing) has a dominant influence across most of the wall-normal range, especially beyond the buffer layer, with Sobol' index values consistently above 0.8. The influence of $d_1$ (streamwise spacing) remains comparatively small with values below 0.2. The multiple term $S_{d_1, d_2}$ is negligible throughout the domain, indicating that the effects of $d_1$ and $d_2$ are largely additive and uncorrelated.
These results clearly demonstrate that spanwise spacing $d_2$ plays a more critical role than streamwise spacing $d_1$ in determining the velocity field and drag reduction characteristics of SHS. Therefore, for robust SHS design under geometric uncertainty, special attention should be given to controlling $d_2$.

\subsection{\label{sec:UQ_TIP}Turbulent intensities profile}
\begin{figure*}
    \centering
    \includegraphics[width=\textwidth]{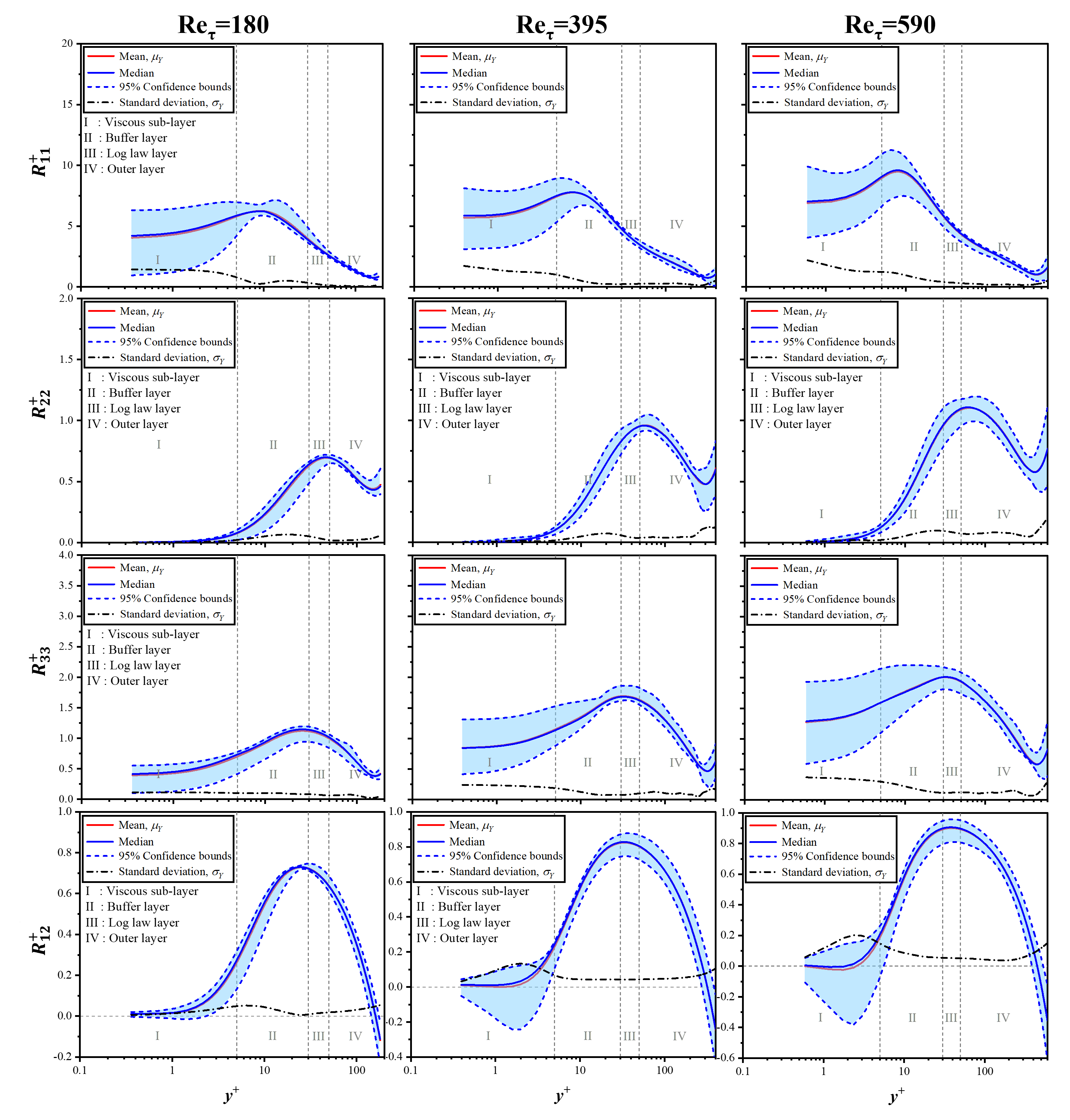}
    \caption{The uncertainty propagation of the Reynolds shear stress($R_{ij}$) profile according to the shear Reynolds number($Re_\tau=180, 395, 590$)}
    \label{fig:UQ_Reynolds_shear_stress}
\end{figure*}
Figure \ref{fig:UQ_Reynolds_shear_stress} shows the uncertainty propagation of the Reynolds stress components $R_{11}^+, R_{22}^+, R_{33}^+, R_{12}^+$ for the three friction Reynolds numbers: $Re_\tau=180, 395,$ and $590$. Each row corresponds to a specific Reynolds stress component, while each column represents a different Reynolds number.
For each plot, the blue solid and dashed lines denote the median and 95\% confidence interval bounds, respectively, while the red line indicates the mean. The shaded region highlights the uncertainty range, and the black dash-dotted line represents the standard deviation $\sigma_Y$.
All results are plotted in wall units as a function of $y^+$, and the boundary layer regions (viscous sublayer, buffer layer, log-law layer, and outer layer) are labeled accordingly. The largest uncertainty is generally observed in the viscous sub-layer, where turbulence intensity is higher and surface-induced effects become more pronounced. As $Re_\tau$ increases, the magnitude of all Reynolds stress components generally increases due to enhanced turbulence levels. This increase is concentrated in the near-wall region, especially within the viscous sublayer and buffer layer.
Among the components, $R_{11}^+, R_{22}^+$, and $R_{33}^+$ all exhibit higher peak values with increasing $Re_\tau$. Notably, $R_{11}^+$ shows the most significant amplification in the viscous sub-layer, indicating intensified streamwise velocity fluctuations in this region at higher Reynolds numbers.

In contrast, the Reynolds shear stress component $R_{12}^+$, which directly contributes to the turbulent momentum transfer, exhibits a reduction near the wall across all $Re_\tau$ cases. This reduction becomes more pronounced with increasing $Re_\tau$, suggesting a stronger drag reduction effect at higher Reynolds numbers. The decrease in $R_{12}^+$ within the viscous sub-layer is indicative of the suppressed near-wall turbulence structures due to the presence of the SHS. Since $R_{12}^+$ represents the dominant contributor to total wall shear stress in turbulent channel flows, this reduction is directly associated with the observed drag reduction. These observations support the conclusion that enhanced slip effects and reduced turbulent momentum transport near the wall are the primary mechanisms underlying the frictional drag reduction achieved by SHSs.  

\begin{figure*}
    \centering
    \includegraphics[width=\textwidth]{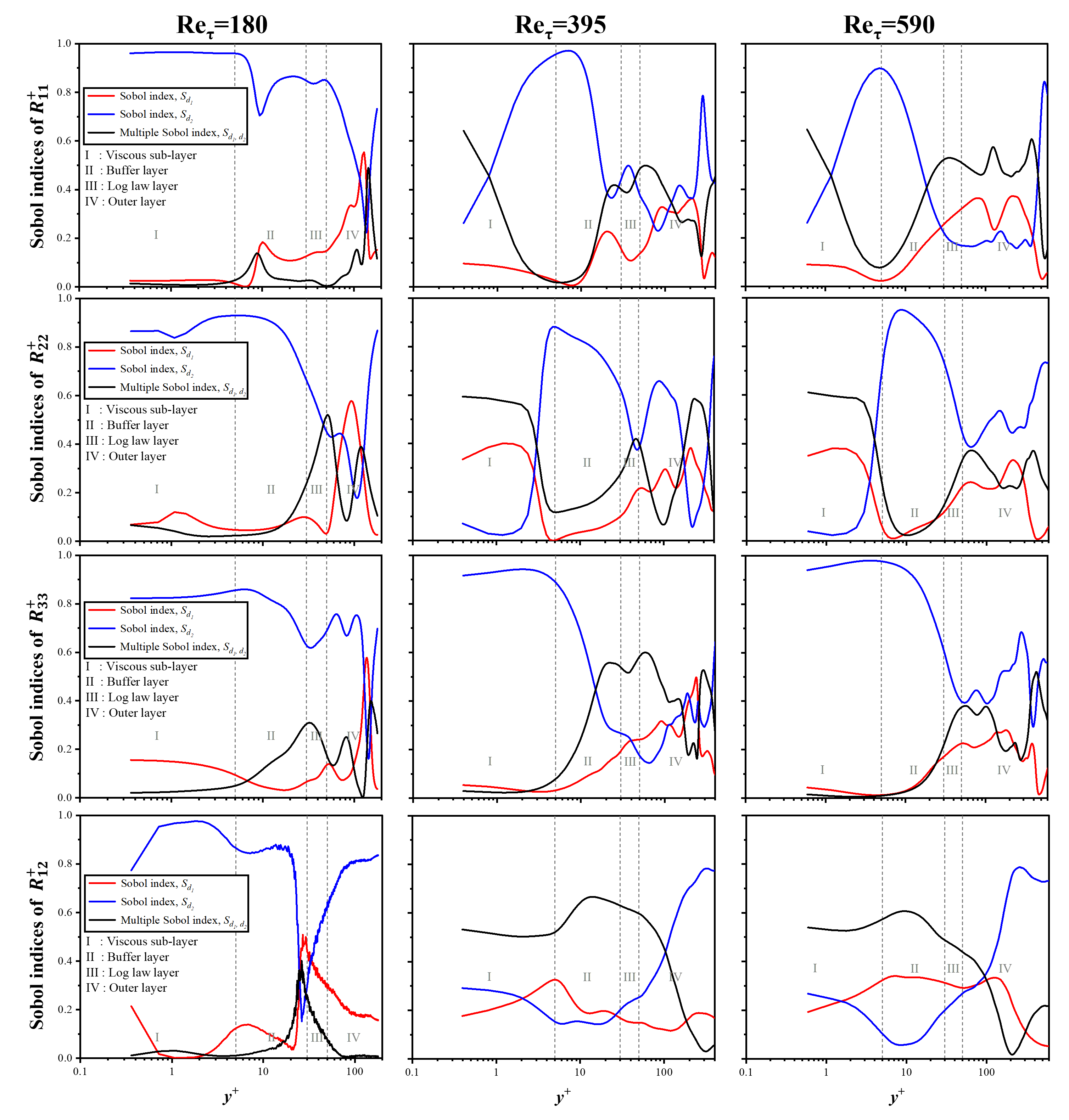}
    \caption{The Sobol' indices of the Reynolds shear stress($R_{ij}$) profile according to the shear Reynolds number($Re_\tau=180, 395, 590$)}
    \label{fig:UQ_Reynolds_shear_stress_Sobol}
\end{figure*}

Figure \ref{fig:UQ_Reynolds_shear_stress_Sobol} shows the Sobol' index profiles of the Reynolds stress, which are similar to the lower panels in Figure \ref{fig:UQ_Velocity_profile}. At $Re_\tau=180$, the Sobol' index of the spanwise spacing ($S_{d_2}$) is dominant across all Reynolds stress components. However, starting from the $Re_\tau=395$, the interaction Sobol' index($S_{d_1, d_2}$) becomes dominant, particularly in the near-wall region. 
In Fig.~\ref{fig:UQ_Reynolds_shear_stress}, the input variable that dominates at the location of maximum uncertainty propagation is considered the most influential for the corresponding Reynolds stress component. The locations of maximum uncertainty propagation are identified as follows: the viscous sub-layer(I) for $R_{11}^+$, the buffer layer(II) for $R_{22}^+$, the viscous sub-layer(I) for $R_{33}^+$, and the viscous sub-layer(I) for $R_{12}^+$. At these respective locations, the dominant input variable is $S_{d_2}$ for $R_{11}^+$, $R_{22}^+$, and $R_{33}^+$. For $R_{12}^+$, $S_{d_2}$ is dominant at $Re_\tau = 180$, whereas the interaction term $S_{d_1,d_2}$ becomes dominant at higher Reynolds numbers.

\subsection{\label{sec:UQ_AIM}Anisotropy invariant map}
\begin{figure*}
    \centering
    \includegraphics[width=\textwidth]{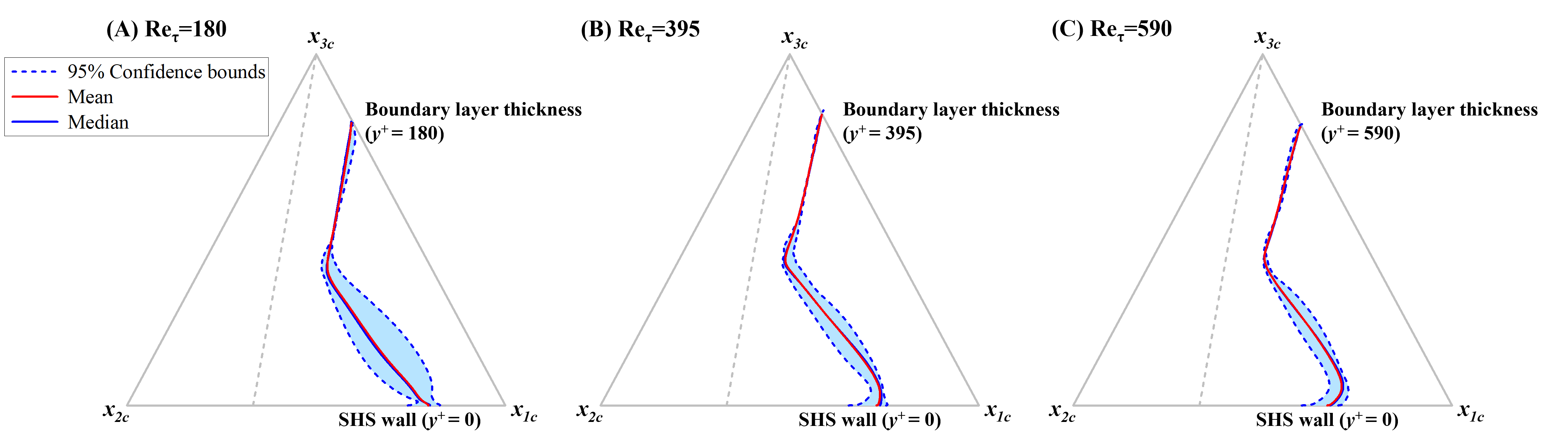}
    \caption{The uncertainty propagation of the barycentric anisotropy invariant map according to the shear Reynolds number($Re_\tau=180, 395,$ and $590$)}
    \label{fig:UQ_AIM}
\end{figure*}
Figure \ref{fig:UQ_AIM} shows the uncertainty propagation on the barycentric AIM for three Reynolds numbers, $Re_\tau = 180$, 395, and 590.
In each subplot, the red line indicates the mean trajectory of the AIM, while the blue line represents the median trajectory. The shaded region bounded by the dashed lines denotes the 95$\%$ confidence interval, illustrating the extent of uncertainty resulting from random input variations in the SHS geometry (streamwise and spanwise spacing).

For all Reynolds numbers, the trajectories begin at the bottom edge of the triangle, which indicates the presence of highly anisotropic turbulence near the SHS wall. As the wall-normal distance increases, the trajectories approach the upper vertex of the triangle, representing isotropic turbulence. The confidence interval is widest near the wall (in the range of $y^+ \lesssim 30$), particularly within the viscous sublayer. This implies that turbulence anisotropy is highly sensitive to SHS geometric variability in the near-wall region, while the outer region is less affected.

As the Reynolds number increases, the uncertainty propagation in the AIM becomes narrower, particularly near the SHS wall. This suggests that the influence of geometric randomness in the SHS configurations on turbulence anisotropy weakens at higher Reynolds numbers. The elevated inertial forces found in high-$Re_\tau$ turbulence likely reduce the sensitivity of the near-wall anisotropy structure to the surface pattern variations.
At lower Reynolds numbers, the mean trajectory in the AIM shows a direct transition from a strong one-component turbulence state near the SHS wall to the isotropic state, bypassing the typical two-component turbulence region. This deviation from canonical turbulence behavior suggests a stronger influence of surface patterning at $Re_\tau=180$. However, as the Reynolds number increases, the trajectory begins to recover the classical path observed in fully-developed channel flows, progressing from two-component turbulence near the wall, through a one-component region, and finally approaching isotropy. This trend is seen most clearly at $Re_\tau=590$.

\section{Conclusion}
This study conducted a unified UQ analysis on turbulent drag reduction over SHSs by systematically parameterizing the streamwise and spanwise spacing of post-type structures. By varying these geometric parameters ($d_1, d_2$), a wide range of SHS patterns, including ridge-type, transverse-ridge-type, and post-type surfaces, were represented and analyzed using high-fidelity DNS with NekRS.

A surrogate model was constructed via PCE to quantify the effects of geometric uncertainties. The results revealed that spanwise spacing ($d_2$) has a significantly greater influence on drag reduction than streamwise spacing ($d_1$), with the Sobol' index of ($d_2$) being up to 14 times higher than that of ($d_1$). This trend was consistently observed across all friction Reynolds numbers studied ($Re_\tau = 180$, $395$, and $590$).

Furthermore, barycentric AIM and Reynolds stress analyses demonstrated that SHSs not only reduce drag but also alter turbulence structures, particularly by weakening coherent vortical structures near the wall. The uncertainty propagation seen in the AIM shows that the effect of SHS pattern variability becomes less significant as the Reynolds number increases, indicating a reduced sensitivity to near-wall turbulence anisotropy at higher inertial regimes.

Overall, the unified UQ was conducted to offer both predictive capability and physical insights into how SHS geometric spacing influences drag reduction performance, providing valuable design guidance for robust SHS configurations under uncertain conditions.

\begin{acknowledgments}
The research was supported by the Agency for Defense Development by the Korean Government (UD230502DD) and through the National Research Foundation (NRF) of Korea funded by the Ministry of Education (NRF-2022R1I1A3063464).
\end{acknowledgments}

\nocite{*}
\bibliographystyle{aipnum4-1}
\bibliography{aipsamp.bib}

\end{document}